\documentstyle[12pt]{article}
\newcommand{\be}{\begin{equation}}
\newcommand{\ee}{\end{equation}}
\newcommand{\bea}{\begin{eqnarray}}
\newcommand{\eea}{\end{eqnarray}}

\begin{document}
\begin{titlepage}
\title{Primordial Fluctuations from Inflation: a Consistent Histories Approach}
\author{David Polarski\\
\hfill \\
Lab. de Math\'ematique et Physique Th\'eorique, EP93 CNRS\\
Universit\'e de Tours, Parc de Grandmont, F-37200 Tours (France)\\
\hfill\\
D\'epartement d'Astrophysique Relativiste et de Cosmologie,\\
Observatoire de Paris-Meudon, 92195 Meudon cedex (France)}

\date{\today}
\maketitle

\begin{abstract}
We show how the quantum-to-classical transition of the cosmological 
fluctuations produced during an inflationary stage can be described using 
the consistent histories approach. We identify the corresponding histories 
in the limit of infinite squeezing. To take the decaying mode into account, 
we propose an extension to coarse-grained histories.
\end{abstract}

\end{titlepage}

\section{Introduction}
In the inflationary paradigm \cite{lin}, all inhomogeneities in the 
universe originated from primordial quantum fluctuations of some scalar 
field(s), the so-called inflaton(s). 
It is then possible to calculate these fluctuations \cite{haw82}, to track 
their evolution untill today and to make predictions, for each given model, 
concerning the formation of large-scale structures in the universe and the 
anisotropy of the Cosmological Microwave Background. The latter will be 
measured with high accuracy till small angular scales (see for example 
\cite{pl})
A comprehensive understanding of the quantum-to-classical 
transition of these fluctuations is required is order to explain the 
obviously classical nature of the inhomogeneities, both in matter and 
radiation, observed on cosmological scales. It was already shown that the 
fluctuations undergo a quantum-to-clasical transition as a result of their 
peculiar dynamics due to the accelerated expansion of the universe during 
the inflationary stage \cite{PS96}. 
This dynamics leads to a very highly squeezed state \cite{Gr,Al94,AA94}, 
where the squeezing 
parameter $|r_k|$ satisfies $|r_k|\gg 1$ . This is expressed in 
the Heisenberg representation by a vanishingly small decaying mode, nowadays 
on cosmological scales, compared to the growing mode. This is why 
inflationary fluctuations that are in principle observable today, after they 
reenter the Hubble radius, have a stochastic amplitude but a fixed phase. 
Note that random amplitude and fixed phase does {\it not} point necessarily 
to a quantum origin, but rather to a primordial origin \cite{PLB95,Al94} of 
the fluctuations. The inflationary paradigm explains elegantly this 
primordial origin. This has observational consequences, for instance the 
existence of secondary acoustic (Sakharov) peaks in the 
small-angle Cosmological Microwave Background anisotropy.
The amplitudes have, in most models where the fluctuations arise from vacuum 
initial states, a Gaussian distribution. However the above results can be 
extended to a non-Gaussian distribution when fluctuations are produced in 
non vacuum initial sates \cite{LPS97}. In other words, this 
quantum-to-classical transition is essentially independent of the initial 
state. The sensitivity of this process to the environment was also 
considered \cite{KPS98} and the 
result was found that for a wide class of interactions this 
transition still holds, independently of the fluctuations initial state at 
the inflationary stage, the amplitude basis defining the pointer basis. 
This quantum-to-classical transition that takes place solely as a result of 
the dynamics is very intriguing. 
The remaining coherence of the initial quantum state can be described to 
very high accuracy with classical {\it stochastic} terms. In the cosmological 
context one is certainly willing to accept the probabilistic description of 
the fluctuations due to an indeterminacy in the initial conditions. 
This fact really becomes non-trivial when the fluctuations are quantized. 
In that case, a deterministic evolution of the wave 
function leads effectively to a classical evolution with stochastic initial 
conditions. The non-relativistic free quantum particle 
at very late times provides yet another surprising example where an analogous 
transition is taking place \cite{KP98}. 
We present here a way to describe this transition using the consistent 
histories approach \cite{CH} (see also \cite{dec} for a review). We feel that 
the consistent histories approach 
is conceptually very appropriate for the description of 
the quantum-to-classical transition taking place for the primordial 
fluctuations of inflationary origin. Furthermore, the Heisenberg picture 
which is used for the description of the (second quantized) fluctuations, 
enters in a natural way in this picture too.
We first briefly review this approach and we show how, for the inflationary 
fluctuations, histories can be defined. 

\vskip 4mm

\section{Decoherent histories}

In this formalism, the evolution of a quantum system is given by specifying 
a set of alternative histories. These are defined by a set 
of Heisenberg projection operators $P^i_{\alpha_i}(t_i)$, at a sequence of 
times $t_i$ with $t_1<t_2<...<t_n$. 
A single history corresponds to 
a particular choice $\alpha_i$ for each time $t_i$, i.e. to a particular set 
of indices $\{\alpha\}\equiv \{ \alpha_1, \alpha_2,..., \alpha_n \}$. The 
Heisenberg projection operators $P^i_{\alpha_i}(t_i)$ satisfy the following 
conditions
\be
P^i_{\alpha_i}(t_i)~P^{i}_{\alpha'_i}(t_i)=P^i_{\alpha_i}(t_i)~
\delta_{\alpha_i \alpha'_i}~\label{excl},
\ee
and
\be
\sum_{\alpha_i}P^i_{\alpha_i}=Id\label{compl}
\ee
Condition (\ref{excl}), which says that the projection operators are 
orthogonal, 
means that the histories $\{\alpha\}$ are exclusive, 
while condition (\ref{compl}), which says that the projection operators 
$P^i_{\alpha_i}(t_i)$ form a complete set at each time $t_i$, means that the 
set of alternative histories is exhaustive. 
We can now define a history operator $C_{\alpha}$
\be
C_{\alpha}\equiv P^n_{\alpha_n}~P^{n-1}_{\alpha_{n-1}}...P^2_{\alpha_2}~
P^1_{\alpha_1}~,
\ee
which gives the branch state vector 
\be
C_{\alpha}~|\Psi \rangle~,
\ee
when applied to the initial state vector $\Psi$, where we assume that we 
start with a pure state (the case of relevance for inflation). 
Sets of alternative histories with vanishing interference are 
said to decohere. In that case, it is possible to attach consistently a 
probability $p(\alpha)$ to each individual history $\{\alpha\}$
\be
p(\alpha)= \langle \Psi|C^{\dagger}_{\alpha}~C_{\alpha}|\Psi \rangle~. 
\ee
Interference is 
mesured by the decoherence functional $D(\alpha~\alpha')$
\be
D(\alpha~\alpha')=\langle \Psi|C^{\dagger}_{\alpha}~C_{\alpha'}|\Psi \rangle~.
\label{funct}
\ee
Hence we have in the case of decoherent histories
\be
D(\alpha~\alpha') = p(\alpha)~\delta_{\alpha \alpha'}~.
\ee
These definitions are straightforewardly generalized when one starts with a 
mixed state and the corresponding density matrix $\rho$. In that case we have 
\be
D(\alpha~\alpha') = {\rm Tr}\left (C_{\alpha}~\rho~C^
{\dagger}_{\alpha'}\right )= p(\alpha)~\delta_{\alpha\alpha'}~.
\ee

\section{Quantum fluctuations from inflation}

Let us consider now the primordial quantum fluctuations produced during 
the inflationary stage. 
Regarding the quantum-to-classical transition, the tensorial fluctuations 
(or gravitational waves) can serve as the paradigm for both types of 
fluctuations, scalar and tensorial, of inflationary origin.
The physics of these quantum fluctuations was already 
studied in depth, we give now the essential results using the notations 
of \cite{PS96}. The quantities $y_{\bf k}$, resp. $p_{\bf k}$, are the Fourier 
tansforms of the amplitude $y$, resp. the conjugate momentum $p$, all these 
quantities being time-dependent. 
The conformal time $\eta\equiv \int^t \frac{dt'}{a(t')}$ is used.

The dynamics of the 
system is particularly transparent in the Heisenberg representation. 
We have, using the amplitude field modes $f_k(\eta)$ with $\Re f_k\equiv
f_{k1}$ and $\Im f_k\equiv f_{k2}$, $f_k(\eta_0)=1/\sqrt{2k}$,
(we adopt a similar notation for all quantities)
\begin{eqnarray}
y(\bf k,~\eta)
&\equiv& f_k(\eta)~a({\bf k},\eta_0)+f_k^*(\eta)~
a^{\dag}(-{\bf k},\eta_0)\nonumber\\
&=&f_{k1}(\eta)~~e_y({\bf k})-f_{k2}(\eta)~~e_p({\bf k})
\label{yk}
\end{eqnarray}
and the momentum field modes $g_k(\eta)$,~ $g_k(\eta_0)=\sqrt{{k\over 2}}$,
\begin{eqnarray}
p(\bf k,~\eta)
&\equiv& -i\bigl\lbrack g_k(\eta)~a({\bf k},\eta_0)
-g_k^*(\eta)~a^{\dag}(-{\bf k},\eta_0)\bigr\rbrack\nonumber\\
&=&g_{k1}(\eta)~~e_p({\bf k})+ g_{k2}(\eta)~~e_y({\bf k}).\label{pk}
\end{eqnarray}
The time independent operators $e_p({\bf k})\equiv 
\sqrt{\frac{2}{k}}~p({\bf k},\eta_0)$, resp. $e_y({\bf k})\equiv 
\sqrt{2k}~y({\bf k},\eta_0) $, satisfy 
\begin{equation}
\langle e_y({\bf k})~e_y^{\dagger}({\bf k}')\rangle=
\langle e_p({\bf k})~e_p^{\dagger}({\bf k}')\rangle=
\delta^{(3)}({\bf k}-{\bf k}')
~~~~~~e_{y,p}^{\dagger}({\bf k})=e_{y,p}(-{\bf k})~.\label{cor} 
\end{equation}
They obey the commutation relations 
\begin{equation}
[e_i({\bf k})~,~e_j^{\dagger}({\bf k}')]=2i~\delta_{ij}~\delta^{(3)}
({\bf k}-{\bf k}')~,~~~~~~i,j=y,p~.\label{com}
\end{equation} 
When the initial state is the vacuum state, all (non vanishing) correlation 
functions are derived from (\ref{cor}). 
The field modes can be parametrized by three parameters, the squeezing 
parameter $r_k$, the squeezing angle $\phi_k$ and the rotation angle 
$\theta_k$. 

If $\lambda$ is the physical wavelength of the perturbation, 
$R_H\equiv \frac{a}{\dot a}$ (in units for which $c=1$) the Hubble radius and 
$a(t)$ the scale factor of the FRW metric then, 
time evolution in the regime $\lambda\gg R_H$
leads to an extreme squeezing which persists when 
the perturbation reenters the Hubble radius, i.e. for $\lambda < R_H$. As a 
result, it is possible to take $f_{k2}\to 0$ and $g_{k1}\to 0$ (see e.g. 
\cite{PS96}). 
The quantum coherence is then expressible in {\it classical stochastic} terms: 
for a given ``realization" $y_{\bf k}$ of the fluctuation field, we  
have for its canonical momentum
 $p_{\bf k}\simeq \frac{g_{k2}}{f_{k1}}~y_{\bf k}\equiv p_{{\bf k},cl}$, 
the {\it classical} momentum for large squeezing, i.e.,
for $|r_k|\to \infty$.
This almost perfect classical correlation is nicely exhibited with the help 
of the Wigner function -- a well-known example of (quasi) probability density 
in phase-space \cite{PS96,LPS97c,KLPS98}. 

We consider now the operators $y_{\bf k}$ separately for each {\bf k} but, 
for brevity of notation, we will drop in the following the subscript 
${\bf k}$ (or k).
In the limit of a perfect classical correlation, the system is described by a 
set of decoherent histories, to each single history a probability can be 
assigned, these probabilities add to one.
The projection operators defining the histories in that case are given by 
\be
P^i_{\alpha_i}(t_i)=| y_i\rangle \langle y_i |~,
\ee
where $y_i$ stands for a particular value of the operator $y(t)$ at 
time $t_i$. We can now consistently assign a probability (density) to 
each individual trajectory $(y(t_n)~y(t_{n-1})$...$y(t_1))$ in amplitude space, 
where all the $y(t_i)$ lie on the classical trajectory 
(for $|r_k|\to \infty$) that goes through $y(t_1)$ at time $t_1$.
Indeed, probability is conserved along classical 
trajectories passing through $y_i$ (at some late time $t_i$), with 
\be
p(\alpha)\equiv |\Psi(y_0,\eta_0)|^2~~~~~y_i=f_{k1}(\eta_i)y_0\equiv 
y_{i,cl}~~~~~~~~~~~~\forall i~.\label{prob} 
\ee
In the limit where the decaying mode is negligible, i.e. $f_{k2}\to 0$, 
(\ref{prob}) holds {\it for all realizations} $y_i$ at time $t_i$ of the 
amplitude. This is the highly non-trivial property of our (isolated) 
{\it pure} state allowing for its description in classical terms. 
In particular, we can consistently define the joint probability 
\be
W(y_n,t_n;y_{n-1},t_{n-1};...;y_2,t_2;y_1,t_1)dy_n...dy_1
\ee
to find the amplitude $y$ between $y_i$ and $y_i +dy_i$ at each late time 
$t_i$. Of course this joint probability is Markovian and satisfies
\bea
W(y_n,t_n;y_{n-1},t_{n-1};...;y_2,t_2;y_1,t_1)=p(y_n,t_n;y_{n-1},t_{n-1})\times\\
p(y_{n-1},t_{n-1};y_{n-2},t_{n-2})...p(y_2,t_2;y_1,t_1){\cal P}(y_1,t_1)~,
\eea
where $p(y_i,t_i;y_{i-1},t_{i-1})$ is the conditional probability density 
to have $y(t_i)=y_i$ provided we had $y(t_{i-1})=y_{i-1}$. Further, 
${\cal P}(y_1,t_1)$ is the probability for $y(t_1)=y_1$ and it satisfies 
\be
{\cal P}(y_1,t_1)=|\Psi(y_0,\eta_0)|^2=p(\alpha)~~~~~
y_1=f_{k1}(\eta_1)~y_0
\ee
Actually, we have very generally 
\be
p(y_i,t_i;y_{j},t_{j})=\delta \left (y(t_i)-\frac{f_1(t_i)}{f_1(t_{j})}~ 
y_{j} \right )~~~~~~1\leq j\leq n\label{delta}
\ee
Equation (\ref{delta}) is valid in the limit that the decaying mode is 
negligible.
The consistent histories picture exhibits nicely the classical 
{\it stochastic} nature of the fluctuations: in the limit $|r_k|\to \infty$ 
one should not expect to find the system on one single classical trajectory. 
Rather, it can be found on any classical trajectory with a certain probability.
Each of these classical trajectories constitutes one of the possible 
alternative histories to which a probability can be assigned.
Usually only probability amplitudes can be 
assigned to trajectories due to quantum interference \cite{Feyn}. While the 
precise physical meaning of the individual trajectories is a subject of 
(endless) debate, the crucial point here is the ability to consistently 
assign a probability to each history (``classical trajectory'').

For large enough, though nevertheless finite squeezing parameter $|r_k|$, 
which is the case relevant for fluctuations of inflationary origin, one may 
wish to take the decaying mode into account. In that case we 
suggest a more realistic description of our system in terms of coarse-grained 
histories corresponding to coarse-grained trajectories in amplitude space. 
Instead of assigning a certain probability to the history 
$(y(t_n)~y(t_{n-1})$...$y(t_1))$, we now assign a probability to the history 
$(\Delta y(t_n)~\Delta y(t_{n-1})$...$\Delta y(t_1))$ where the projection 
operator $P^i_{\alpha_i}(t_i)$ is now given by
\be
P^i_{\alpha_i}(t_i)=| \Delta y_i\rangle \langle \Delta y_i |~,\label{proj}
\ee
where $|\Delta y_i\rangle$ corresponds to a state for which the amplitude 
$y(t)$ at time $t_i$ is localized within the interval $\Delta y_i$. 
The corresponding decoherence functional $D_{\Delta}(\alpha~\alpha')$ can 
then be written in the following suggestive way
\be
D_{\Delta}(\alpha~\alpha') = \int_{\alpha}~{\cal D}y~
\int_{\alpha'}~{\cal D}y'~D(y(t),y'(t))~,\label{int}
\ee 
where $D(y(t),y(t'))$ corresponds to the histories constructed with the 
projection operators $| y(t)\rangle \langle y(t) |$ 
and $\{\alpha \}\equiv (\Delta y_n,...,\Delta y_1)$. 
Decoherence is achieved when 
\be
D_{\Delta}(\alpha~\alpha')\approx 0~.\label{deco1}
\ee
We can think of a particular history as a series of slits trough which our 
system has to pass at the successive times $t_i$, the condition (\ref{deco1}) 
then expresses the almost complete absence of diffraction. This almost 
complete absence of diffraction was shown with a slit (thought) experiment 
in \cite{KLPS98}.
Note that the integrals appearing in (\ref{int}) are really functional 
integrals, integrals over all the possible paths \cite{har} that can belong to a given 
coarse-grained history $\{\alpha \}$, not 
just classical paths. In 
inflation we will have that $\Delta y_i$ is tremendously small, far beyond our 
observational capabilities. This is where the smallness of the decaying mode 
enters.   

To summarize, we have shown how the quantum-to-classical transition of 
quantum fluctuations of inflationary origin can be nicely described in the 
consistent histories picture. We have identified the histories and the 
corresponding projection operators in the limit $|r_k|\to \infty$ and we have 
extended our description, using coarse grained histories, to the more 
realistic case where the decaying mode, though tremendously tiny compared 
to the growing mode, is nevertheless present.
In the case of fluctuations arising from inflation, the decaying mode is 
definitely negligible compared to the growing mode so that the limit 
$|r_k|\to \infty$ safely applies for most purposes. However, depending on the 
level of precision with which one is willing, or has, to describe the 
fluctuations one may wish to take the decaying mode into account. 
We propose to use coarse-grained histories for this purpose. In a sense, it 
amounts to describe the evolution 
of our system in terms of classical trajectories with a very tiny quantum 
noise which is due to that piece of the amplitude (``position'') operator 
connected to the decaying mode.
It is also possible to consider decoherence in the system-environment couple 
in the consistent histories picture \cite{fin} and we leave these points for 
further investigation.
The consistent histories picture is known to be very fruitfull in many 
problems related to decoherence in particular, and to the interpretation of 
quantum mechanics in general. It is therefore gratifying that this approach 
can be used in the context of inflation as well. In particular all the 
interpretational questions arising in this picture can be considered for the 
particular, but all-important, case of fluctuations of inflationary origin.

%\section*{Acknowledgements}


\begin{thebibliography}{99}

\bibitem{lin}A.~Linde, Rep. Prog. Phys. 47 (1984) 925; 
 Particle physics and inflationary cosmology (Harwood, New York,
 1990); E.~Kolb, M.~Turner, The Early Universe (Addison-Wesley,
 Redwood City, 1990).
\bibitem{haw82} S.W. Hawking, Phys. Lett. B 115 (1982) 295;
 A.A. Starobinsky, Phys. Lett. B 117 (1982) 175;
 A.H. Guth and S-Y. Pi, Phys. Rev. Lett. 49 (1982) 1110.
\bibitem{pl} http://astro.estec.esa.nl/SA-general/Projects/Planck/
 http://map.gsfc.nasa.gov/
\bibitem{PS96} D. Polarski and A.A. Starobinsky,
 Class. Quantum Grav. 13 (1996) 377.
\bibitem{Gr} L.P. Grishchuk and Y.V. Sidorov, Phys. Rev. D 42 (1990) 3413.
\bibitem{Al94} A. Albrecht, P. Ferreira, M. Joyce, and T. Prokopec,
 Phys. Rev. D 50 (1994) 4807. 
\bibitem{AA94}A. Albrecht, Report hep-th/9402062.
\bibitem{PLB95}D.~Polarski and A.A.~Starobinsky, Phys. Lett. B 356 (1995) 196.
\bibitem{LPS97} J. Lesgourgues, D. Polarski, and A.A. Starobinsky,
 Nucl. Phys. B 497 (1997) 479.
\bibitem{KPS98} C. Kiefer, D. Polarski, and A.A. Starobinsky, 
 Int. Journ. Mod. Phys.~D 7 (1998) 455.
\bibitem{KP98} Kiefer C and Polarski D {\em Report gr-qc/9805014}, 
 {\it Ann. Physik} 7 (1998) 137.
\bibitem{CH} R. Griffiths, J. Stat. Phys. 36 (1984) 219.\\
 R. Griffiths, Phys. Rev. A 54 (1996) 2759. \\
 R. Omn\`es, J. Stat. Phys. 53 (1988) 893.\\
 R. Omn\`es, The interpretation of Quantum Mechanics (Princeton 
 University Press, Princeton 1994).\\
 M. Gell-Mann and J. Hartle, Complexity, Entropy and the Physics of 
 Information, SFI Studies in the Sciences of Complexity, Vol.VIII, 
 (Edited by W. Zurek) (Addison Wesley, Reading 1990), p.425.
\bibitem{dec} Giulini D, Joos E, Kiefer C, Kupsch J,
 Stamatescu I O and Zeh H D, Decoherence and the Appearance
 of a Classical World in Quantum Theory (Springer 1996).
\bibitem{LPS97c} J. Lesgourgues, D. Polarski, and A.A. Starobinsky,
 Class. Quantum Grav. 14 (1997) 881.
\bibitem{KLPS98} C. Kiefer, J. Lesgourgues, D. Polarski, and A.A. 
 Starobinsky, Class. Quantum Grav. 15 (1998) L67.
\bibitem{Feyn} Feynman R P and Hibbs A R, Quantum Mechanics
 and Path Integrals (McGraw-Hill 1965). 
\bibitem{har} J. Hartle Phys. Rev. D 10 (1991) 3173.
\bibitem{fin} J. Finkelstein Phys. Rev. D 47 (1993) 5430.
\end{thebibliography}
\end{document}